\documentclass[lettersize,journal]{IEEEtran}
\usepackage[hidelinks, urlcolor=black, colorlinks=true, linkcolor=blue, citecolor=blue]{hyperref}
\usepackage[caption=false,font=footnotesize]{subfig}
\usepackage{array}
\usepackage{cite}
\usepackage{amsmath,amssymb,amsfonts}
\usepackage{siunitx}
\usepackage{pifont}
\usepackage{algorithmic}
\usepackage{graphicx}
\usepackage{booktabs}
\usepackage{makecell}
\usepackage{textcomp}
\usepackage{xcolor}
\usepackage{orcidlink}
\usepackage{acronym}\acrodef{3GPP}[3GPP]{3rd generation partnership project}
\acrodef{AI}[AI]{artificial intelligence}
\acrodef{WI}[WI]{Wireless InSite\textsuperscript{\textregistered}}
\acrodef{IP}[IP]{Internet Protocol}
\acrodef{MSE}[MSE]{mean square error}
\acrodef{B5G}[B5G]{beyond 5G}
\acrodef{KPI}[KPI]{key performance indicator}
\acrodef{5G}[5G]{5th generation}
\acrodef{SLA}[SLA]{service-level agreement}
\acrodef{MAPE}[MAPE]{Mean Absolute Percentage Error}
\acrodef{5G/6G}[5G/6G]{5th and 6th generation}
\acrodef{QoS}[QoS]{quality of service}
\acrodef{LoRaWAN}[LoRaWAN]{long range wide area network}
\acrodef{LPWAN}[LPWAN]{low power wide area network}
\acrodef{IoT}[IoT]{internet of things}
\acrodef{RT}[RT]{ray tracing}
\acrodef{DT}[DT]{digital twin}
\acrodef{NDT}[NDT]{network digital twin}
\acrodef{ITU}[ITU]{International Telecommunication Union}
\acrodef{MLP}[MLP]{multilayer perceptron}
\acrodef{UDP}[UDP]{User Datagram Protocol}
\acrodef{TCP}[TCP]{Transmission Control Protocol}
\acrodef{PDR}[PDR]{packet delivery ratio}
\acrodef{SF}[SF]{spreading factor}
\acrodef{ED}[ED]{end-device}
\acrodef{GW}[GW]{gateway}
\acrodef{UAV}[UAV]{unmanned aerial vehicle}
\acrodef{CDF}[CDF]{cumulative distribution function}
\acrodef{MILP}[MILP]{mixed integer linear programming}
\acrodef{BILP}[BILP]{binary integer linear programming}
\acrodef{MBNLP}[MBNLP]{mixed binary non-linear programming}
\acrodef{MNLP}[MNLP]{mixed non-linear programming}
\acrodef{MWF}[MWF]{multi-wall and multi-floor}
\acrodef{ILP}[ILP]{integer linear programming problem}
\acrodef{GA}[GA]{genetic algorithm}
\acrodef{RMSE}[RMSE]{root mean squared error}
\acrodef{EPSO}[EPSO]{evolutionary particle swarm optimization}

\def\BibTeX{{\rm B\kern-.05em{\sc i\kern-.025em b}\kern-.08em
    T\kern-.1667em\lower.7ex\hbox{E}\kern-.125emX}}
\begin{document}
\title{LoRaWAN Gateway Placement for Network Planning Using Ray Tracing-based Channel Models}
\author{\IEEEauthorblockN{Cláudio Modesto\raise0.5ex\hbox{\orcidlink{0009-0005-5184-8030}}, Lucas Mozart\raise0.5ex\hbox{\orcidlink{0009-0000-2092-4399
}}, Glauco Gonçalves\raise0.5ex\hbox{\orcidlink{0000-0003-1341-5339
}}, Cleverson Nahum\raise0.5ex\hbox{\orcidlink{0000-0001-9644-5394
}}, \\Bruno Castro\raise0.5ex\hbox{\orcidlink{0000-0003-4601-3205
}},
and Aldebaro Klautau\raise0.5ex\hbox{\orcidlink{0000-0001-7773-2080
}},~\IEEEmembership{Senior Member,~IEEE}}
%\IEEEauthorblockA{\IEEEauthorrefmark{1}LASSE, Federal University of Pará, Brazil}
%\IEEEauthorblockA{Email: \url{claudio.barata@itec.ufpa.br}}
}% <-this % stops an unwanted space

%\markboth{To be submitted on IEEE/ACM Transactions on Networking}%
%{}
\maketitle

\begin{abstract}
Network planning for long range wide area networks (LoRaWAN) relies heavily on the channel models used to estimate wireless coverage and connectivity. Consequently, the quality of gateway (GW) deployment decisions may be strongly affected by the propagation assumptions adopted during the planning process. Given this motivation, this work investigates how different channel models influence the placement of LoRaWAN GWs, formulating an optimization problem that contrasts stochastic and empirical models with ray-tracing-based models. To this end, we developed a framework that integrates ray tracing (RT) simulators with a discrete-event network simulator. Using this framework to generate LoRaWAN data metrics, we employ an optimization model that determines the optimal GW placement under different channel models, received power constraints, and network scenarios. Our results show that the optimized solution is highly sensitive to the chosen channel model, even when considering the same scenarios with different RT simulators, revealing a clear trade-off between computational cost and the fidelity of the solution to real-world conditions.
\end{abstract}


\maketitle

\section{INTRODUCTION}
\IEEEPARstart{N}{etwork} planning is a fundamental task in wireless communication systems, particularly in \ac{LPWAN} deployments, due to the high variability and inherent challenges of these networks, including very low bitrates, limited throughput, and stringent energy constraints~\cite{lima2024lora, pagano2024application, Loh2022, said2025}. Furthermore, network planning is a crucial task concerning the economic viability of the system, as it enables strict control over hardware investments and ongoing maintenance costs~\cite{lima2024lora, Matni2020, Costa2024}. In this way, it ensures that the infrastructure meets requirements such as \ac{QoS} and reliability for specific application scenarios, preventing connectivity failures and guaranteeing stable data delivery~\cite{Pires2024}. Therefore, this kind of task is also important for scalability, as it prepares the network to support future growth in traffic load and device density without performance degradation~\cite{Loh2021}. Regarding the types of \acp{LPWAN}, one of the most popular is \ac{LoRaWAN}, which is ideally suited for \ac{IoT} scenarios that require long-range communication, low power consumption, and low cost. %Currently, a large number of \ac{IoT} devices are in operation, and this figure is expected to grow substantially in the coming years. In this context, some studies forecast that there will be approximately $40$ billion connected devices worldwide by $2030$~\cite{sinha2024state}. %In this context, one of the main \ac{LPWAN} technologies currently used to connect \ac{IoT} devices is the previously mentioned \ac{LoRaWAN} network. 

\ac{LoRaWAN} is a protocol based on the LoRa modulation scheme~\cite{alliance2017lorawan, silva2021} and operates using three main components: \ac{ED}, \acf{GW}, and a network server. Regarding these components, one of the most important aspects of \ac{LoRaWAN} network planning is the position and number of \acp{GW}~\cite{pagano2024application} as these factors directly affect the achievable coverage area and the ability of \acp{ED} to meet their communication requirements. Poor network planning may result in \acp{ED} experiencing connectivity issues or insufficient link~\cite{LohICC2023}.

To determine the optimal \ac{GW} configuration for a given network topology, measurement-based approaches~\cite{lima2024lora, cruz2022} and simulation-based methods can be employed to plan the best positions and the number of \acp{GW} intended to fulfill the \ac{ED} requirements. In this work, we assume a simulation-based approach, taking into account different channel models. In this sense, common models include stochastic channels, such as those defined by 3GPP~\cite{3gppTR38901}, and empirical approaches, such as log-distance models~\cite{Correia2023, Pires2024} and Okumura-Hata~\cite{Matni2020, Loh2022}, which can offer flexibility in terms of parameters and low computational cost. To perform site-independent simulations (which do not depend on a specific 3D scenario), we must accept a tradeoff between the fidelity of the results in terms of large scale parameters, such as the coverage pattern of each transmitter. Thus, to achieve reliable performance in this task, accurately simulating the wireless channel is indispensable for ensuring the quality of wireless simulations and, consequently, the reliability of network planning results~\cite{ruz20233d}. Such accuracy can be attained using specialized tools, such as \ac{RT} simulators~\cite{borges2024, modesto2025}.

Nonetheless, the use of site-specific channel models in \ac{LoRaWAN} \ac{GW} placement problems remains underexplored in the literature. To address this gap, we propose a framework that integrates \ac{RT} and discrete-event simulators.\footnote{Source code is available at: \url{https://github.com/lasseufpa/rt-lorawan}} This integrated environment enables the evaluation and comparison of different network scenarios in terms of channel configuration for the \ac{GW} placement problem. The results from the proposed experiments show a relevant impact in terms of the number of \acp{GW} and the position of each one in the optimized solution when a site-specific channel is considered instead of the commonly used site-independent channels. Numerically, in a more detailed 3D scenario (detailed in terms of the number of faces), the experiment with the site-specific channel from Sionna\textsuperscript{\texttrademark} \ac{RT} and \ac{WI} using the X3D algorithm~\cite{wirelessInsite} indicates that 9 and 3 \acp{GW}, respectively, should be deployed in a grid with 100 possible positions while also considering power constraints with a fixed threshold. In contrast, using standard models such as COST-231 and 3GPP-UMa, the suggested numbers of \acp{GW} are 11 and 1 \acp{GW}, respectively. Therefore, since site-specific channel generation inherently incurs a higher computational cost (mainly when a large 3D scenario is used)~\cite{modesto2025, Zhu2024, testolina2024}, this discrepancy highlights a trade-off between solution accuracy and the computational cost that is ultimately determined by the chosen channel model. 
 
Therefore, our main contributions to this paper are:

\begin{itemize}
    \item A simulation framework for LoRaWAN scenarios that integrates commercial and open-source \ac{RT} simulators alongside a discrete-event network simulator.
    \item An evaluation of the \ac{GW} placement optimization problem using different levels of channel fidelity, considering both site-specific and site-independent approaches.
    \item An analysis of the computational costs, considering different channel models for network planning.
\end{itemize}

The remainder of this paper is organized as follows: Section~\ref{sec:related_work} reviews the related literature, highlighting the main characteristics of existing works concerning their optimization approaches, types of channels, and main objectives. In Section~\ref{sec:gw_placement}, we describe the proposed framework, which integrates \ac{RT} simulators and a discrete-event network simulator used in a \ac{GW} placement optimization problem. For this problem, we detail in this section the decision variables, constraints, and the objective function addressed. Section~\ref{sec:experiments} describes the network setup and the methodology used to evaluate the impact of channel models on the \ac{GW} placement optimization. The results from this methodology are discussed in Section~\ref{sec:results}. Finally, Section~\ref{sec:conclusion} summarizes the main contributions of this work and discusses directions for future research.

\section{RELATED WORK}
\label{sec:related_work}
The \ac{GW} placement problem in \ac{LoRaWAN} networks has been extensively investigated in prior research~\cite{Loh2021, Matni2020, Pires2024, Loh2023, Correia2023}. In particular, Pires et al.~\cite{Pires2024} addressed \ac{GW} placement optimization in \ac{UAV}-assisted scenarios. Their study formulates the task as a \ac{MILP} problem, incorporating multiple constraints related to slice-specific \ac{QoS} requirements, transmission power limits, and the number of devices assigned to each \ac{SF}. The network evaluation relies on an empirical channel model based on the log-distance path-loss formulation. 
Correia et al.~\cite{Correia2023} also employ the log-distance path loss model, but with the stochastic component of shadowing to plan LoRaWAN networks for smart agriculture, focusing on a large fruit-growing area in Brazil. The main contribution of the paper is a comparison of four clustering algorithms: K-Means, MiniBatch K-Means, Bisecting K-Means, and Fuzzy C-Means, to optimize \ac{GW} placement and quantity. The overall performance is evaluated using the uplink delivery rate and a stochastic energy consumption model. Another study that utilizes a stochastic propagation model is presented by Prudêncio et al.~\cite{prudencio2025}. The authors employ the \ac{MWF} channel model, which is stochastic due to its shadowing component. The research focuses on applying a \ac{GA} for network planning of a \ac{LoRaWAN} \ac{GW} architecture within a multi-floor indoor environment. %The primary objective is to minimize the distance between \ac{LoRaWAN} devices and the gateway to improve network connectivity and reduce energy consumption.

\newcommand{\xmark}{\ding{55}}%
\newcommand{\cmark}{\ding{51}}%

\begin{table*}[!h]
\caption{Works related to the \ac{GW} placement problem}
\label{tab:gw_positioning_rw}
\centering
\scalebox{1}{\begin{tabular}{lccccc}
\toprule
  
\textbf{Work} & \textbf{Channel model} & \makecell{\textbf{Solution}\\\textbf{method}}   & \makecell{\textbf{Site-dependency}} & \textbf{\makecell{\textbf{Main objective}  }}  \\
\midrule
%\makecell{Hussain et al., 2017~\cite{Hussain2017}} &~\cite{degli_2004} & \xmark  & \cmark \\ 
Matni et al., 2020~\cite{Matni2020} & Okumura-Hata & Heuristic  & Independent & Minimize CAPEX and OPEX\\ 
\midrule
Correia et al., 2023~\cite{Correia2023} & Log-distance & Heuristic &  Independent & Comparing clustering algorithms\\
\midrule
Loh et al., 2022~\cite{Loh2022} & Okumura-Hata & Heuristic & Independent & Minimize message collisions\\
\midrule
Cruz et al., 2022~\cite{cruz2022} & UFPA model & Metaheuristic & Independent &  \makecell{Minimize the number of \acp{GW}} \\
\midrule
Loh et al., 2023 
\cite{Loh2023} & Okumura-Hata & Heuristic & Independent & Minimize message collisions\\
\midrule
Pagano et al., 2024~\cite{pagano2024application} & Unspecified & Heuristic & Independent & Minimize energy consumption\\
%\midrule
%Lima et al., 2024~\cite{lima2024lora} & --- & --- & -- & --- & --\\
\midrule
Mhatre et al., 2024~\cite{mahtre2024} & Unspecified & \makecell{Heuristic and \\metaheuristic} & Independent & Minimize the number of \acp{GW}\\
\midrule
Pires et al., 2024~\cite{Pires2024} & Log-distance & Solver  & Independent & \makecell{Minimize the number of \acp{GW}\\ and the number of \ac{ED} with the same \ac{SF}}\\ 
\midrule
Prudêncio et al., 2025~\cite{prudencio2025} & \makecell{\acs{MWF}} & Metaheuristic & Independent & Minimize distance between nodes\\ \midrule
Said et al., 2025~\cite{said2025} & Unspecified & Clustering algorithms & Independent & Maximize the number of covered \acp{ED}\\
\midrule

%\midrule
Our work & From \ac{RT} & Solver & Dependent & Minimize the number of \acp{GW}\\ 
% (maybe power or something like that) requirement
\bottomrule
\end{tabular}}
\end{table*}

Loh et al.~\cite{Loh2022, Loh2023} proposed methods in different works to perform \ac{GW} placement using graph-based heuristics that focus on reducing message collisions and data loss. By mitigating collisions in a network with random channel access and limited retransmissions, these methods improve energy efficiency and extend the battery lifetime of \ac{IoT} devices, which is a crucial issue when discussing this type of scenario. In these works, to define a \ac{GW} transmission range to be used in the \ac{GW} placement, the authors employed an empirical channel model, namely the Okumura-Hata model. Another work that uses the same propagation model is described in Matni et al.~\cite{Matni2020}. In this work, the authors introduce DPLACE, a \ac{GW} placement framework tailored to dynamic \ac{IoT} environments. It divides the \ac{IoT} devices into groups and addresses each group with specific requirements. The main optimization objective is to maximize the \ac{PDR}, while ensuring that both capital expenditure (CAPEX) and operational expenditure (OPEX) remain within acceptable bounds. Following the empirical approaches, Cruz et al.~\cite{cruz2022} introduced a new methodology for \ac{LoRaWAN} planning in a complex Amazonian scenario on the Campus of the Federal University of Pará (UFPA) in Brazil, composed of forests and buildings. The authors proposed significant updates to the regional UFPA propagation model, where the model was adjusted using a \ac{GA} to minimize the \ac{RMSE} between predicted path loss and field measurements. The core objective of the study is to employ an \ac{EPSO} metaheuristic within a multi-objective optimization. This strategy aims to achieve a balance between maximizing the coverage area and minimizing the number of \acp{GW} to reduce deployment costs.

In Pagano et al., \cite{pagano2024application}, the authors propose a method similar to the aforementioned works, based on graph theory but within the specific context of massive \ac{IoT} water distribution systems. Their work introduces an application-aware approach that utilizes the application weight centrality (AWC) metric in the optimization process for \ac{GW} placement. Therefore, instead of relying solely on network connectivity, the AWC incorporates hydraulic pressure values as weights for the graph nodes, prioritizing \ac{GW} positioning near areas with critical pressure variations. Mhatre et al.~\cite{mahtre2024} introduced two distinct algorithms for \ac{GW} placement, primarily aimed at enhancing energy efficiency and reliability in dense networks. The first algorithm uses the analysis of connected components (CC-Place), while the second is based on the metaheuristic simulated annealing (SA-Place). The authors also compare their results with another graph-based heuristic and a clustering approach based on fuzzy C-Means, demonstrating that their methods provide superior performance primarily by reducing message collisions. Said et al.~\cite{said2025}, using an unspecified site-independent channel model with the FLoRa framework~\cite{Slabicki2018}, evaluate different clustering algorithms for \ac{GW} placement with the objective of maximizing the number of covered \acp{ED}. Their results show that clustering methods such as K-means can yield efficient \ac{GW} placement solutions, complementing traditional approaches based on exact solvers, heuristics, and metaheuristics.

\begin{figure*}
    \centering
    \includegraphics[scale=0.5]{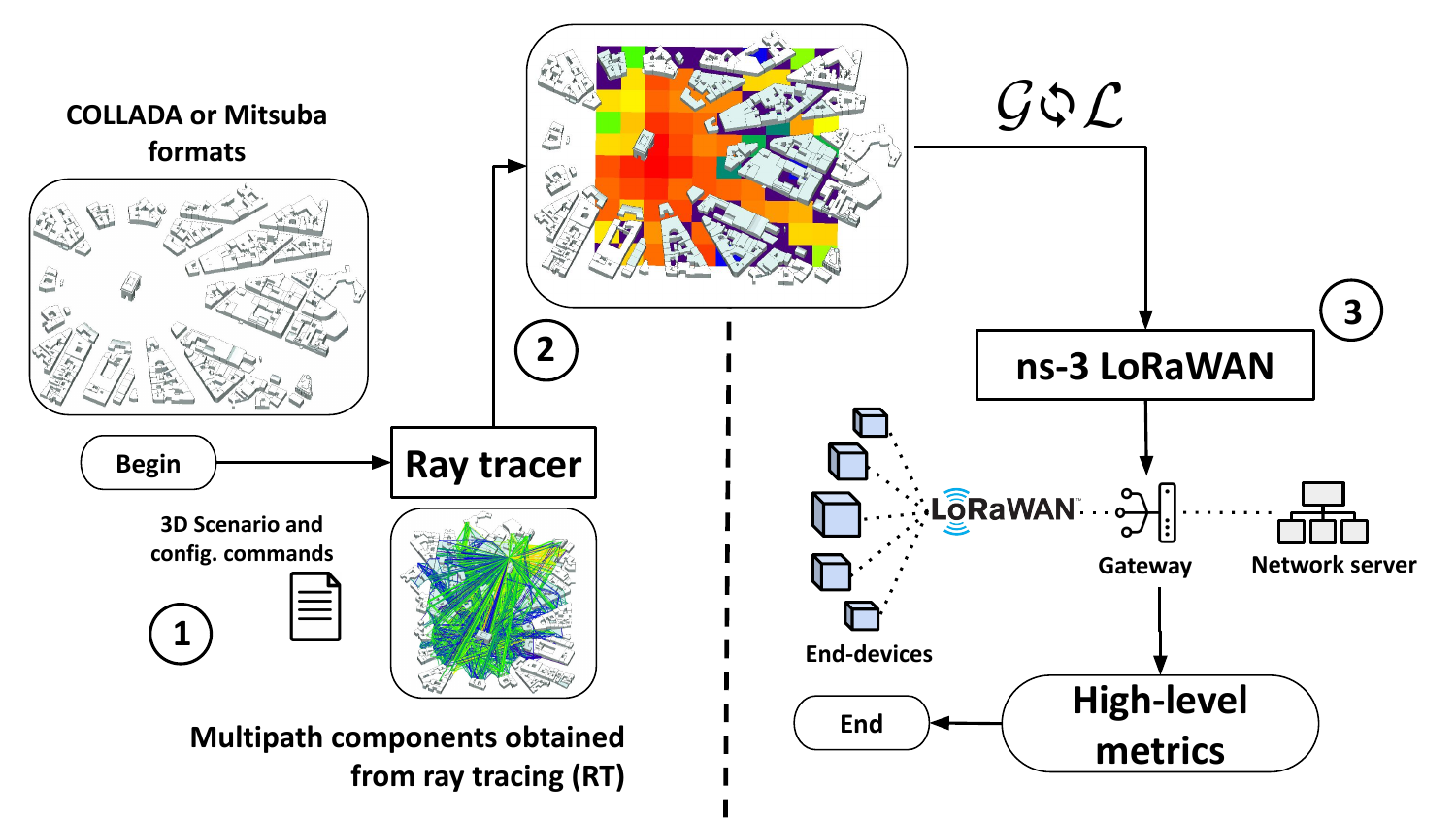}
    \caption{Proposed framework integrating a \ac{RT} and discrete-event network simulator used to simulate \ac{LoRaWAN} scenarios in different network layers.}
    \label{fig:rt_ns3_lorawan}
\end{figure*}

Finally, the work by Ruz-Nieto et al. \cite{ruz20233d} presents a simulation framework with a network simulator, a $3$D engine, and a \ac{RT} tool that realistically models the performance of \ac{LoRaWAN} networks. They conducted their experiments using a deterministic channel-modeling approach based on \ac{RT} and compared the results with empirical and stochastic models, highlighting the benefits of a more precise and accurate representation of radio-wave propagation. Although that work does not address network planning in \ac{LoRaWAN} using site-specific channel models, it provides important comparative analysis of modeling approaches. Moreover, it does not consider the algorithmic optimization of \ac{GW} placement, which remains a distinct research gap that our work addresses.

All the previously mentioned related works that focus on \ac{GW} placement rely on stochastic or empirical approaches in their simulations, which use site-independent channel models and do not adequately explore site-specific ones. These site-specific models are the focus of our work, given their importance for accurate network planning for a specific deployment
scenario. Therefore, Table~\ref{tab:gw_positioning_rw} compares these works in terms of the channel model used (including its dependence on the deployment site) and the solution method adopted to address the problem, with respect to their main objectives.

\section{RT-LoRaWAN FRAMEWORK FOR GATEWAY PLACEMENT OPTIMIZATION}

\label{sec:gw_placement}
In this section, we characterize the proposed framework by analyzing the input–output relationships of its individual modules. We then demonstrate its primary application to the \ac{GW} placement optimization problem, where the objective is to minimize the number of deployed \acp{GW}.

\subsection{RT-LoRaWAN framework}
The integration between \ac{RT} simulators and discrete-event network simulators, such as ns-3, strongly depends on the target communication system and the high-level performance metrics of interest. A recent example of this type of integration is the use of \ac{RT}-generated channels to enable more realistic \ac{5G} simulations, for instance, by coupling the 5G-LENA~\cite{cttc5GLENAModule} module with channel impulse responses produced by Sionna \ac{RT}~\cite{albuquerque2024}. A similar approach can be adopted for 802.11 Wi-Fi systems, using the same \ac{RT} engine to provide detailed channel models for the network simulator~\cite{Zubow2026, Pegurri2025}. However, for \ac{LoRaWAN} simulations, to the best of our knowledge, this type of integration has not yet been explored in the literature. Since it is essential to simulate and visualize a communication system at different levels of abstraction, we address this gap with a framework that integrates \ac{RT} simulators (both open-source and commercial) with a discrete network simulator (ns-3), aiming to provide \ac{LoRaWAN} simulations with different levels of fidelity regarding wireless channels. 

The operational flow of this proposed framework is defined by Fig.~\ref{fig:rt_ns3_lorawan}. This figure demonstrates that our framework’s pipeline operates. In particular, it shows that the framework can function both in conjunction with \ac{RT} and ns3-\ac{LoRaWAN} simulators, as well as as a standalone system (e.g., using only the ns-3 \ac{LoRaWAN}). If conducted jointly using the realistic channel from \ac{RT}, the pipeline starts with (1) defining the 3D scenarios that should be used and the \ac{RT} configuration parameters. For 3D scenarios, the required input format depends on the selected \ac{RT} simulator. If Sionna \ac{RT} is used, the scenario must be provided in the Mitsuba format~\cite{Mitsuba3}. If the \ac{WI} is used instead, the 3D scenario must follow the Digital Asset Exchange (DAE) format (COLLADA). Finally, to obtain these 3D scenarios, it is possible to import from OpenStreetMap, as proposed in \cite{albuquerque2024}, noting that after the import, it is necessary to convert to the aforementioned file formats using software such as Blender~\cite{blenderBlenderFree}.

\begin{figure}[!h]
\centering

\subfloat[Path gain coverage map evaluated using \ac{WI} with X3D \ac{RT} algorithm.]{
\includegraphics[scale=0.2]{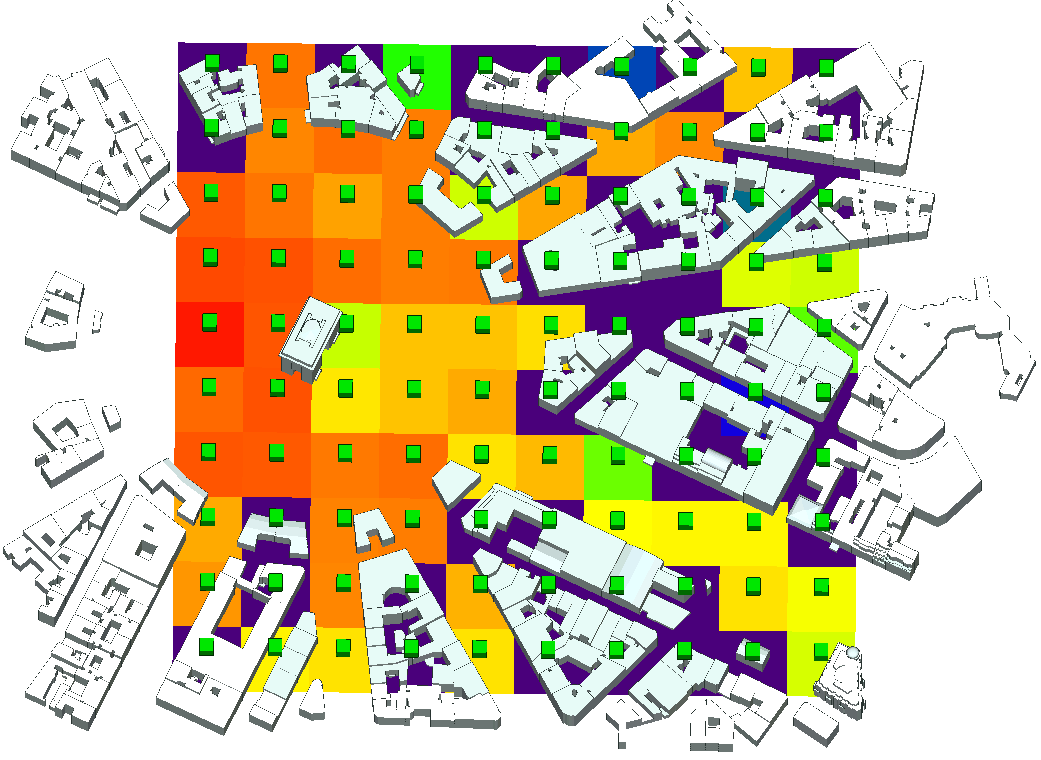}
\label{fig:cm_wi}
}
\hfill
\subfloat[Path gain coverage map evaluated using Sionna \ac{RT}.]{
\includegraphics[scale=0.25]{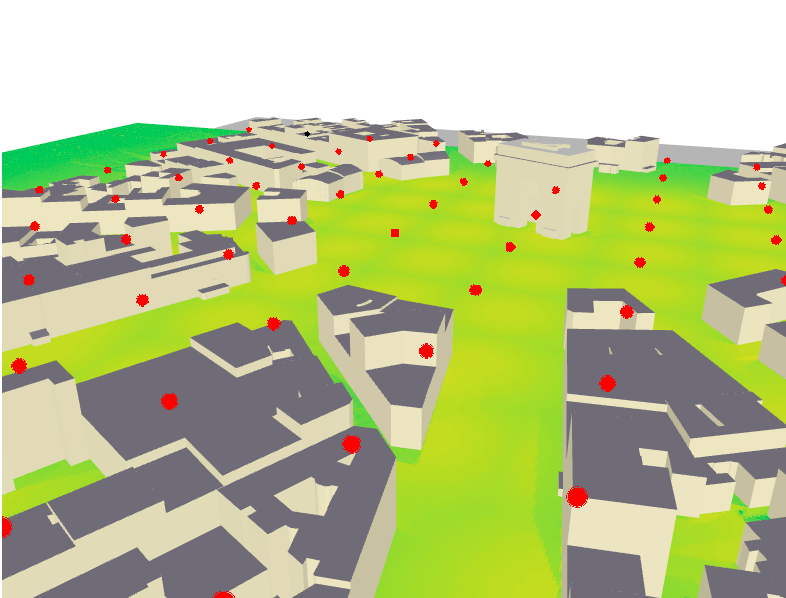}
\label{fig:cm_sionna}
}
\caption{Example of graphical ray tracer output evaluated using \ac{WI} and Sionna \ac{RT}.}
\label{fig:ray_tracers}

\end{figure}

Regarding the parameters configuration, the framework includes the transmitter position, the carrier frequency used, the radiation pattern, polarization, the \ac{RT} algorithms, and other specific parameters for the \ac{RT} simulation. It is important to note that at this simulation level, we abstract the concept of a \ac{GW} or \ac{ED}; in the physical layer, we consider only transmitters and receivers~\cite{ruz20233d}. After completing a \ac{RT} simulation in step (2), we can extract several physical-layer features. One of these features is the path gain $\mathcal{G}$ in \si{dBm} for each \ac{ED}, which is saved in separate files per \ac{GW} position. In this sense, this metric is extracted from a matrix that represents the path gain within a specific area portion, the size of which is defined in the previous step of the \ac{RT} simulation. This information will serve as the interface between both simulators. Figs. \ref{fig:cm_wi} and \ref{fig:cm_sionna} show the path gain calculated using \ac{WI} and Sionna \ac{RT}, respectively. Armed with this information, we use it in the ns-3 \ac{LoRaWAN} module to configure the \ac{LoRaWAN} channels. However, since ns-3 requires the path loss $\mathcal{L}$ in \si{dBm}, we convert the path gain into path loss, noting that $\mathcal{L} = -\mathcal{G}.$

Finally, with the path loss information in hand, this can be used in step (3) to simulate \ac{LoRaWAN} network scenarios from a high-level perspective, in the sense of evaluating and obtaining packet-level metrics such as the \ac{PDR}, without explicit collision modeling, which is the ratio of packets received to packets sent. Moreover, in this level of simulation, we have greater flexibility to consider energy configuration models and different channel models for comparison purposes. For example, at this level of simulation, we can define parameters such as the simulation time and the \ac{SF}.

\subsection{Gateway placement optimization}
We can use our proposed framework to support data collection in the context of a mono-objective \ac{GW} placement problem that assumes a grid with a given number of positions, where multiple positions must be selected to deploy a \ac{GW} to provide coverage for all remaining positions, which act as \ac{ED}. Furthermore, all the terms used in this work are defined in Table~\ref{tab:paper_terms}.

\begin{table}[!h]
    \centering
    \caption{Description of terms used}
    \begin{tabular}{l|l}
    \toprule
        Term & Description \\
    \midrule
         $\mathcal{D}$ & Set of \ac{ED} \\
         $\mathcal{P}$ & Set of positions \\
         $\mathcal{K}$ & Set of chosen \acp{GW} \\
         $D$ & Total number of \ac{ED} \\
         $P$ & Total number of positions \\
         $K$ & Total number of chosen \acp{GW} \\
         $d$ & $d$-th \ac{ED} \\
         $p$ & $p$-th \ac{GW} position \\
         $k$ & $k$-th chosen \ac{GW} position \\
         $\rho$ & Power threshold \\
         $\alpha$ & Received power \\
         $\beta$ & Coverage indicator \\
         $x$ & Decision variable \\
         $\text{PDR}_{\text{no\_collision}}$ & \ac{PDR} without collision \\
         $\text{PDR}_{\text{collision}}$ & \ac{PDR} with collision \\
         $f(\cdot)$ & Function that maps a position to an  $\text{PDR}_{\text{no\_collision}}$\\
    \bottomrule
    \end{tabular}
    \label{tab:paper_terms}
\end{table}

%\begin{equation}
%    \mathrm{minimize} \sum_{p\in \mathcal{P}} %x^{p}.
%\label{eq: objective_function}
%\end{equation}

We can formally classify this problem as a \ac{BILP} with a binary decision variable $x^{p}$ that identifies whether a \ac{GW} exists at a given position. The position $p$ belongs to the set of candidate positions $\mathcal{P} = \{1,2,\dots,p,\dots,P\}$,
where $P$ is the total number of possible positions defined over a spatial grid. Each index represents a Cartesian coordinate in the deployment area. Each deployed \ac{GW} in the set $\mathcal{K} = \{1, 2, \dots, k, \dots, K\}$, where $K$ denotes the total number of selected \acp{GW}, must provide coverage to the \acp{ED} in the network and is associated with an aggregated $\ac{PDR}_{\text{no\_collision}}$ such that:
\begin{equation}
\text{PDR}_{\text{no\_collision}}
=
[f(1), f(2), \dots, f(k), \dots, f(K)],
\end{equation}
where $f : \mathcal{K} \rightarrow [0,1]$ is a function that maps a deployed \ac{GW} position to its corresponding aggregate $\ac{PDR}_{\text{no\_collision}}$, and $f(k)$ denotes the aggregated $\text{PDR}_{\text{no\_collision}}$ obtained when a \ac{GW} is placed at position $k$. Moreover, we assume that an \ac{ED} $d$ comes from a set $\mathcal{D} = \{1, 2, \dots, d, \dots, D\}$ of position indices (also representing Cartesian coordinates), consisting of a total of $D$ devices. Therefore, our goal is to minimize the number of deployed \ac{GW} from the set $\mathcal{K}$, leading to the following objective function:
\begin{equation}
\min_{x^{p}} \sum_{p \in \mathcal{P}} x^{p}.
\label{eq: objective_function}
\end{equation}
\begin{figure*}[!h]
\centering

\subfloat[\ac{ED} grid configuration 1.]{
\includegraphics[scale=0.46]{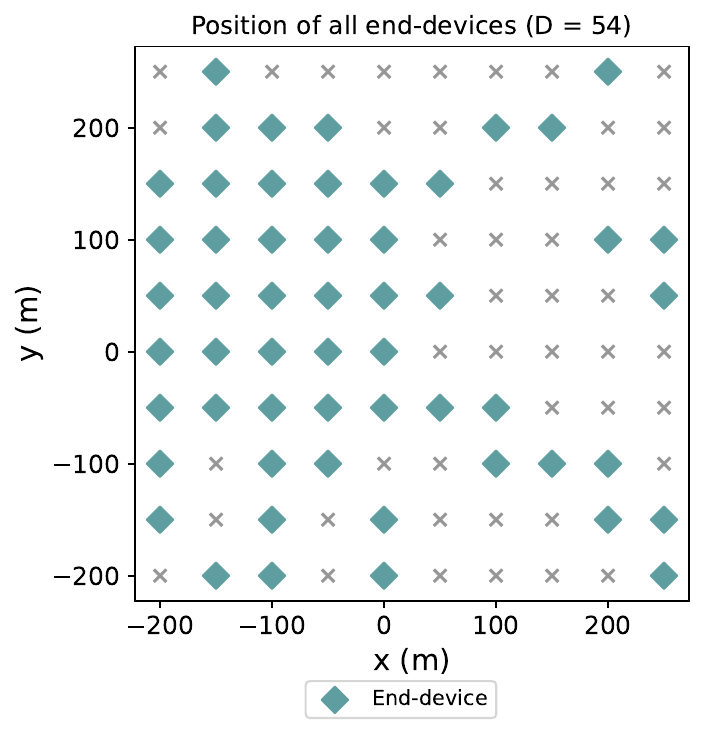}
\label{fig:grid_1}
}
\subfloat[\ac{ED} grid configuration 2.]{
\includegraphics[scale=0.46]{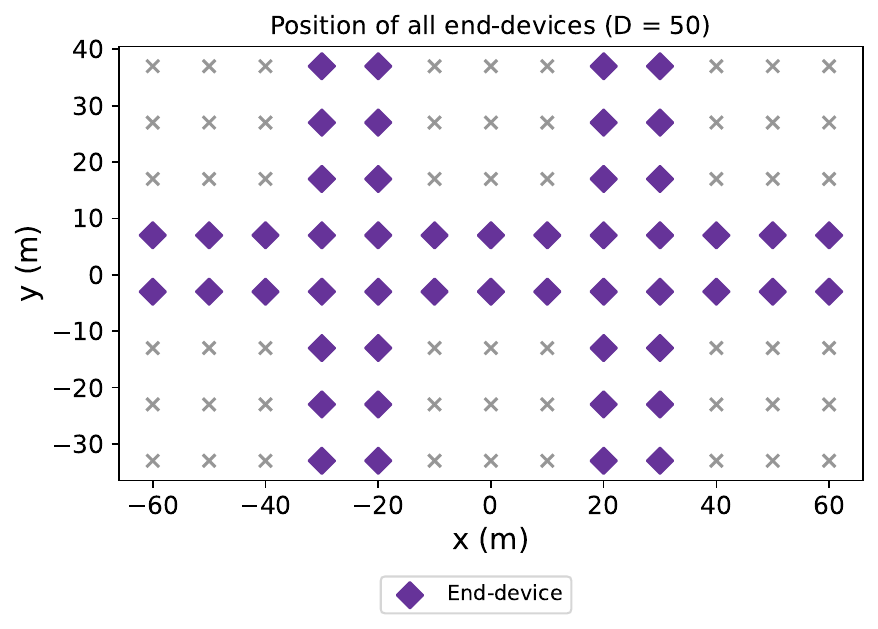}
\label{fig:grid_2}
}
\caption{Spatial \ac{ED} organization that should be covered by a set of GWs at $P$ possible positions.}
\label{fig:example}
\end{figure*}

Regarding the constraints, the objective function defined in Eq.~(\ref{eq: objective_function}) is subject to a received power constraint at the \ac{ED}. In this work, we assume that the received power $\alpha$ is equal to the path gain ($\alpha = \mathcal{G}$), since the transmit power is set to zero \si{dBm} in the data provided by our framework. Thereby, this constraint ensures that communication between an \ac{ED} and a \ac{GW} occurs only when the received power $\alpha$ at the \ac{ED} exceeds a given threshold $\rho$ in \si{dBm}. We assume that we have precomputed whether the $d$-th \ac{ED} is covered by any \ac{GW}. In this sense, this computation assumes that we can calculate the received power for each \ac{ED} once a \ac{GW} position is defined by using the \ac{RT} simulator or through the simulation with site-independent channels using ns-3. Thus, for every possible \ac{GW} location, we can determine in advance whether a given \ac{ED} is covered. With this information, we define the variable $\beta^p_d$, which indicates the coverage of the $d$-th \ac{ED} by the $p$-th \ac{GW}, where $\beta_d^p = 0$ indicates that the $d$-th \ac{ED} is not covered ($\alpha < \rho$) by the \ac{GW} in the $p$-th position, and $\beta_d^p = 1$ indicates that the \ac{ED} is covered ($\alpha \ge \rho$) by the \ac{GW} at the $p$-th position. With this precomputed \ac{ED} coverage information, we then define a set of multiple coverage constraints as follows:
\begin{equation}
    \sum_{p\in \mathcal{P}} \beta_d^p x^{p} \geq 1, \quad \forall d \in \mathcal{D}.
    \label{eq: power_constraint}
\end{equation}

With this standard binary optimization model, we aim to highlight the impact of the channel model on network planning outcomes
under controlled and reproducible conditions, with the objective of demonstrating in the next section, devoted to the proposed experiments, that even with only a small set of constraints, the optimal solution can change abruptly in terms of the number of \acp{GW}, the position of each \ac{GW}, and, as a consequence, negatively overestimate the performance of high-level metrics such as \ac{PDR}.

\section{EXPERIMENTS}
\label{sec:experiments}
To assess how different channel models affect the solutions for the formulated \ac{GW} placement, both in terms of the number of suggested \ac{GW} and the computational cost required to obtain them, we conducted a series of evaluations using our proposed framework. For these evaluations, we considered two network topologies based on uniform grid layouts. In grid configurations 1 and 2, adjacent candidate positions are spaced 50\,\si{m} and 10\,\si{m} apart, respectively. Following a methodology similar to~\cite{Pires2024}, these grids provide $P = 100$ and $P = 104$ candidate \ac{GW} locations. For each candidate position, a \ac{GW} is deployed at a height of 30\,\si{m} in grid configuration 1 and 60\,\si{m} in grid configuration 2. Regarding the organization of the \acp{ED}, we assume that each deployed \ac{GW} serves $D = 54$ devices in grid configuration 1 and $D = 50$ devices in grid configuration 2. Each device is set at a height of $1.4$~\si{m} and is spatially distributed according to the grids shown in Figs.~\ref{fig:grid_1} and \ref{fig:grid_2}. Each valid position is defined to avoid intersecting with any buildings when used in a 3D scenario. With this configuration, for each position in the grid, we place a \ac{GW} to evaluate the received power at the set of \acp{ED}. After computing these received powers for every possible \ac{GW} position, we used them to solve the optimization problem. In summary, the topological characteristics used in each scenario are defined in Table~\ref{tab:topology_char}. 

\begin{table}[!htp]
    \centering
    \caption{Grid topology characteristics for different scenarios}
    \begin{tabular}{ccccc}
    \toprule
       Grid  & $P$ & $D$ & \ac{GW} height (\si{m}) & \ac{ED} height (\si{m})\\
    \midrule
       1  & 100 & 54 & 30 & 1.4\\ 
       2  & 104 & 50 & 60 & 1.4\\ 
    \bottomrule
    \end{tabular}
    \label{tab:topology_char}
\end{table}

The data collected from these proposed topologies was then used in experiments involving two groups of channel models. The first group of channels contains all site-independent models represented by Okumura-Hata, COST-231~\cite{Heath2018}, log-distance, and 3GPP-UMa. Regarding the model parameters, some require specific configurations. The log-distance model was configured with a path-loss exponent of $3.76$ and a reference distance of $32$ \si{m}. For the frequency-dependent models (excluding the log-distance model), we used a carrier frequency of 1 \si{GHz}. Finally, for the models that account for the heights of the \ac{GW} and \ac{ED}, we used the previously stated values of 30 and 60 \si{m} for the \ac{GW} (for each different scenario) and 1.4 \si{m} for the \ac{ED}. With all these parameters configured, the received power is calculated considering the use of only the ns-3 simulator in step (3), depicted in Fig.~\ref{fig:rt_ns3_lorawan}.

The second group of channels is represented by the channel models that come from Sionna \ac{RT} and \ac{WI}. For this second group, we also evaluated the received power using our proposed framework. In this way, following the pipeline shown in Fig.~\ref{fig:rt_ns3_lorawan}, we started the process of collecting the path gain in step (1), using 3D scenarios with different levels of detail from Sionna, called Etoile and Street Canyon~\cite{sionna_rt_docs}, with totals of 13\,058 and 37 faces, respectively. The reason for choosing both scenarios is to evaluate the impact of 3D objects such as buildings, with fewer numbers (Street Canyon scenario) and in a more crowded area with more buildings (Etoile scenario). We used the software Blender to convert the file format to that required by our framework. In these 3D scenarios, we used the same network grid topologies as in the evaluation with site-independent channels to allocate transmitters and receivers (which will be interpreted further by ns-3 as \ac{ED} and \ac{GW}, respectively). In this sense, for the Etoile and Street Canyon scenarios, we used the \ac{ED} grid patterns 1 and 2, respectively. 

To obtain the physical layer metrics, specifically the received power, we calculated it for each \ac{GW} possible position. The parameters used in both ray tracers to calculate the coverage map are described in Table~\ref{tab:rt_parameters}. For the carrier frequency, we assumed a proxy value of $1$ \si{GHz} because the radio-material models in the \ac{RT} simulators do not support properties such as conductivity and permittivity for frequencies below $1$ \si{GHz}. These simulators use the property values provided by \ac{ITU}-R recommendation P.2040-3~\cite{itu2040} for materials such as concrete, brick, glass, and so on. Moreover, since the same material properties can be applied to frequencies below $1$ \si{GHz} as well~\cite{itu2040}, this choice is acceptable. Quantitatively, considering the frequency dependence of free-space path loss, the resulting deviation is relatively small for common \ac{LoRaWAN} frequency bands. For example, the free-space path loss difference between 868 \si{MHz} and 1 \si{GHz} is approximately 1 \si{dB}, while the difference between 915 \si{MHz} and 1 \si{GHz} is approximately 0.8 \si{dB}.

% however, the same electromagnetic properties of the radio materials are defined by the \ac{ITU}-R P.2040-3 recommendation~\cite{itu2040} for the range of GHz. 

%Hence, since the Etoile scenario does not consider any type of ground materials, such as wet ground, medium dry ground, and very dry ground, the values of conductivity and permittivity remain acceptable for frequencies close to 1 GHz~\cite{itu2040}.

\begin{table}[htp]
    \centering
    \caption{\ac{RT} simulation parameters used for the experiments with site-specific channels using \ac{WI} and Sionna \ac{RT}}
    \scalebox{1.1}{\begin{tabular}{lc}
    \toprule
      Simulation parameter & Value\\
    \midrule
       Carrier frequency (\si{GHz})  & $1$ \\
       Radiation pattern & Isotropic \\
       Tx polarization & Vertical \\
       Maximum number of interactions & 6\\
       Tx waveform & Sinusoid\\
       Interactions & Diffraction and reflection \\
\bottomrule
    \end{tabular}}
\label{tab:rt_parameters}
\end{table}

% Graphically, the grid organization and an example of the coverage path gain calculated for a given transmitter (black dot) are shown in Fig.~\ref{fig:coverage_map_example}.

% \begin{figure}[!h]
%     \centering
%     \includegraphics[scale=0.42]{figures/fig_2.pdf}
%     \caption{Grid organization and coverage map example for a given transmitter (black dot) in the Etoile scenario. The red dots represent other possible for \ac{GW} placement.}
%     \label{fig:coverage_map_example}
% \end{figure}

We conducted these simulations considering the \ac{RT}-\ac{LoRaWAN} framework in a configured environment, as defined by Table~\ref{tab:configuration_setup}. In this environment, we primarily used tools such as Sionna \ac{RT}, \ac{WI}, and ns-3~\cite{henderson2008network}. For Sionna \ac{RT} and ns-3 simulations, we utilized a server equipped with an Intel\textsuperscript{\textregistered} Xeon Silver 4514Y, a NVIDIA\textsuperscript{\textregistered} A2 GPU, and 512 \si{GB} of RAM. For \ac{RT} simulations with \ac{WI}, we used a server equipped with an Intel\textsuperscript{\textregistered} Core\textsuperscript{\texttrademark} i7-10700F CPU @ 2.90 GHz, a NVIDIA\textsuperscript{\textregistered} GeForce 3060 GPU, and 64 \si{GB} of RAM. With this setup, we obtained the received power metrics for the two channel groups, which were then used as input to the optimization model implemented in Pyomo~\cite{hart2017pyomo} and solved with GLPK~\cite{gnuGLPKProject}, using the \textit{branch-and-bound} algorithm.

Using the collected data from both groups of channels, we propose four types of evaluations in the two scenarios. The first evaluates the impact of the channel models on \ac{GW} placement with a fixed minimum received power threshold of $\rho = -90$\,\si{dBm}. The second uses the same setup but studies this impact over a range of received power thresholds, varying from $-120$ to $-80$\,\si{dBm}. The third evaluation examines the influence of GW placement under different channel models on high-level performance metrics such as \ac{PDR}. In this evaluation, we assume an ALOHA-like \ac{PDR} model~\cite{Heusse2023}, such that the \ac{PDR}, considering a set of positioned \acp{GW}, is defined as:
\begin{equation}
\text{PDR}_{\text{collision}} =
\dfrac{e^{-2L}}{|\mathcal{K}|}
\sum_{k \in \mathcal{K}}
\text{PDR}^{k}_{\mathrm{no\_collision}},
\label{eq:multigw_pdr}
\end{equation}
\noindent where 
\(\text{PDR}^{k}_{\mathrm{no\_collision}}\) represents the isolated
\ac{PDR} associated with \ac{GW} $k$, obtained from ns-3 simulations
without explicit collision modeling. $L$ denotes the offered traffic load, which is evaluated considering the packet rate $\lambda$ during an airtime $T_{\text{airtime}}$, in such a way that $L=\lambda T_{\text{airtime}}$. In this model, to avoid overly optimistic macro-diversity gains caused by assuming
statistically independent \ac{GW} receptions~\cite{Abakar2022}, we evaluate the collision
impact using the average \ac{GW} $\ac{PDR}_{\text{no\_collision}}$ penalized by the ALOHA contention
factor. Finally, the fourth set of evaluations assesses the computational cost, under different channel models, required to obtain the metrics for the \ac{GW} placement problem.

\begin{table}[!h]
    \centering
    \caption{Setup configuration used to generate data and solve the optimization problem}
    \scalebox{1.1}{\begin{tabular}{lc}
    \toprule
    Tool  & Version \\
    \midrule

    Operating system & Ubuntu 20.04.6 and Windows 10 \\
    Pyomo & 6.9.5   \\
    ns-3 & 3.45 \\
    Sionna \ac{RT} & 1.2.1\\
    GLPK & 4.65\\
    Wireless InSite & 3.4.5 \\
    Blender & 3.6.23\\
    \bottomrule
    \end{tabular}}
    \label{tab:configuration_setup}
\end{table}

\section{RESULTS}
\label{sec:results}
\subsection{Coverage and gateway placement analysis}
\begin{figure}[!h]
    \centering
    \subfloat[Chosen \acp{GW} using grid pattern 1.]{
    \includegraphics[scale=0.55]{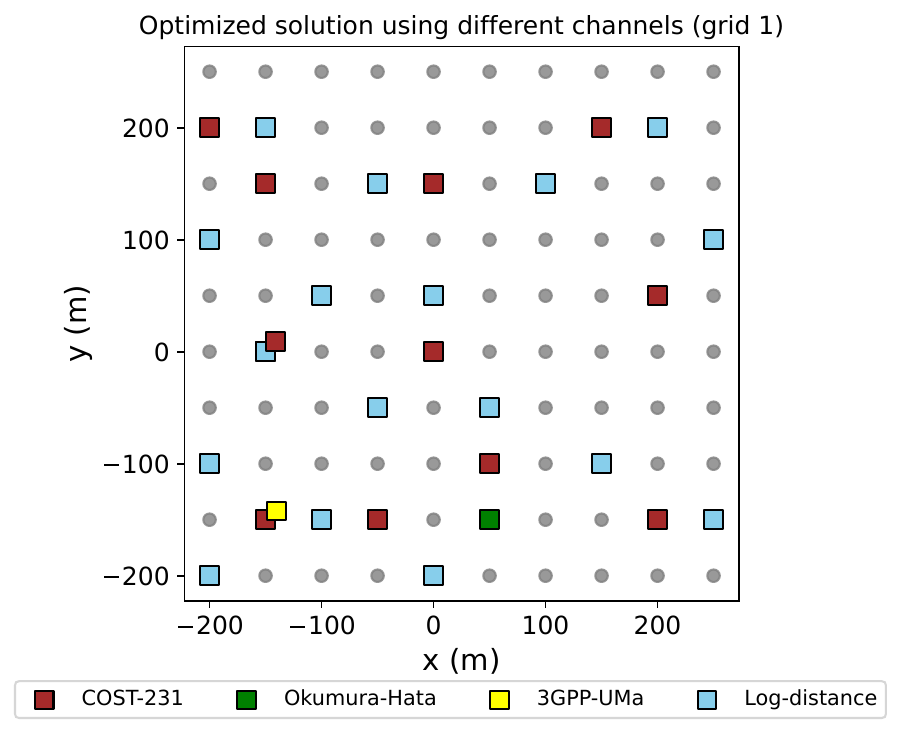}
    \label{fig:site_independent_results_grid1}
    }

    \subfloat[Chosen \acp{GW} using grid pattern 2.]{
    \includegraphics[scale=0.52]{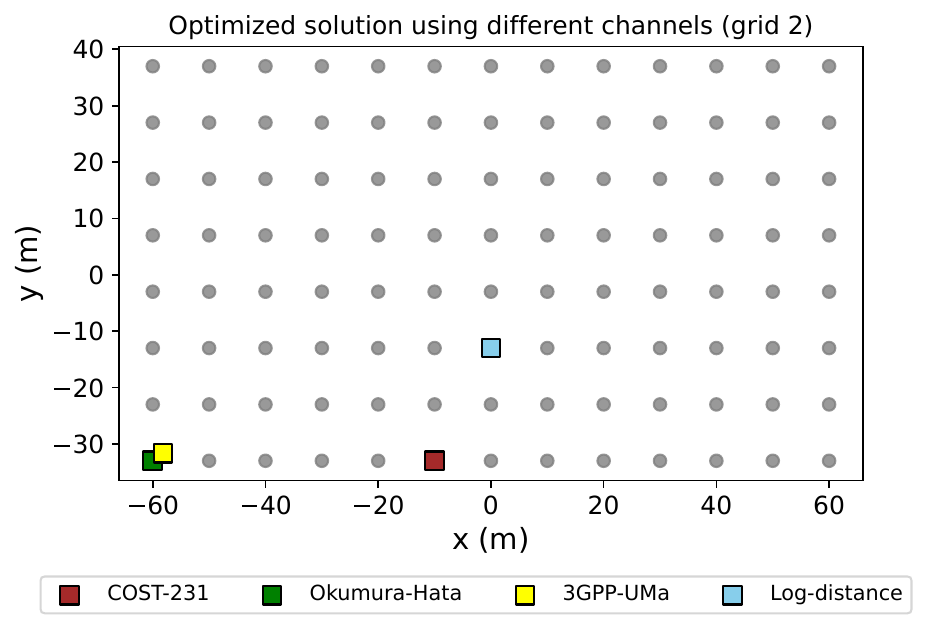}
    \label{fig:site_independent_results_grid2}
    }
    
    \caption{Optimized solution for the \ac{GW} placement using different types of site-independent channels, in this case: COST-231, Okumura-Hata, 3GPP-UMa, and log-distance.}
    \label{fig:site_independent_results}
\end{figure}

In the first set of evaluation, we considered the spatial analysis of the positioned \acp{GW} for the different channels used. To graphically represent this type of experimental result, we focus to show the graphical results with configuration with network grid 1 and 2. Using these patterns, the suggested \acp{GW} in grid 1 and grid 2 are depicted in Figs.~\ref{fig:site_independent_results_grid1} and \ref{fig:site_independent_results_grid2}, respectively, in terms of the position and the number of \acp{GW} allocated. In these figures, we are considering only the site-independent channel models.

In the network organization of grid 1, the optimized solutions are 
strongly influenced by the selected channel model. Using the Okumura–Hata model, the optimization yields a deployment with 1 \acp{GW}, whereas the 3GPP-UMa and COST-231 models result in solutions requiring 1 and 11 \acp{GW}, respectively. In contrast to these models, the log-distance model exhibits markedly different behavior, suggesting 17 \acp{GW}. Finally, because the log-distance model yields a larger number of positioned \acp{GW}, some \ac{GW} locations overlap and are shared by the solutions obtained with the COST-231 channel model. For grid 2, because it represents the simplest scenario, the positional variability of each \ac{GW} is lower than in grid 1. In this case, the simulations for each channel model indicate that a single \ac{GW} is sufficient to meet the requirements.

\begin{table*}[!h]
    \centering
    \caption{Number of suggested \acp{GW}, standard deviation, and the average \ac{ED} received power (in dBm) for each channel model across different grid patterns}
    \scalebox{1.05}{
    \begin{tabular}{lcccccc}
    \toprule
    & \multicolumn{3}{c}{Grid 1} & \multicolumn{3}{c}{Grid 2} \\
    \cmidrule(lr){2-4} \cmidrule(lr){5-7}
    Channel model 
    & \# GW & Avg. ED power & Std. ED power  
    & \# GW & Avg. ED power & Std. ED power \\
    \midrule
    Okumura-Hata  & $1$  & $-78.65$ & $7.47$ & $1$ & $-53.62$ & $6.07$ \\
    
    COST-231 & $11$ & $-72.76$ & $29.34$ & 1 & $-78.14$ & $7.03$ \\
    Log-distance & $17$ & $-68.15$ & $27.32$ & 1 & $-79.85$ & $6.96$ \\
    3GPP-UMa      & $1$  & $-71.31$ & $8.94$ & $1$ & $-60.29$ & $6.16$ \\
    Sionna        & $9$  & $-81.47$ & 3.76 & $1$ & $-83.16$ & $1.89$ \\
    WI (X3D)      & $3$  & $-73.39$ & $6.23$ & $2$ & $-72.73$ & $6.07$ \\
    WI (Full 3D)  & $2$  & $-75.94$ & $7.20$ & $2$ & $-68.93$ & $1.52$ \\
    \bottomrule
    \end{tabular}}
    \label{tab:power_threshold_scenarios}
\end{table*}
 
\begin{figure}[!h]
    \centering
    \subfloat[Chosen \acp{GW} in Etoile scenario.]{
    \includegraphics[scale=0.55]{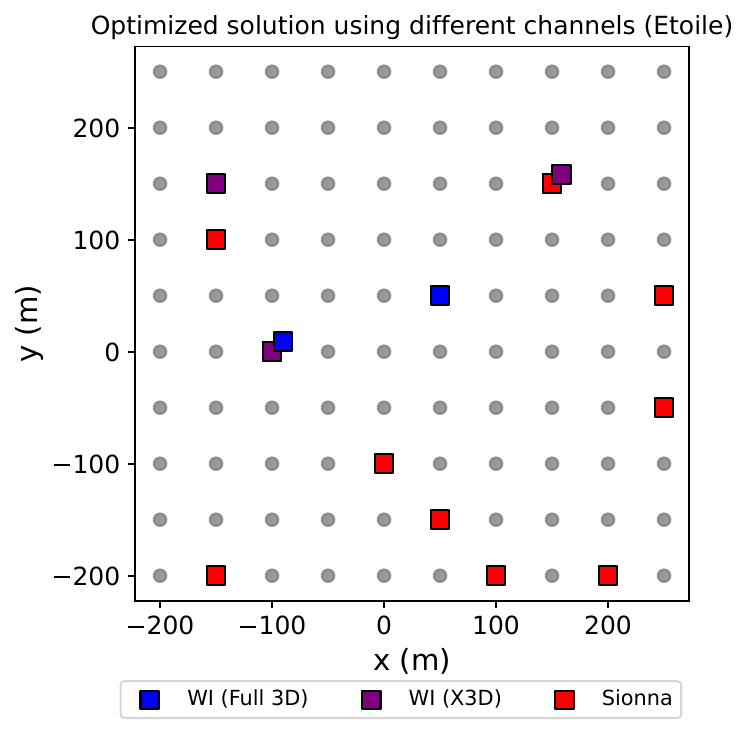}
    \label{fig:site_specific_results_grid1}
    }
    
    \subfloat[Chosen \acp{GW} in Street Canyon scenario.]{
    \includegraphics[scale=0.52]{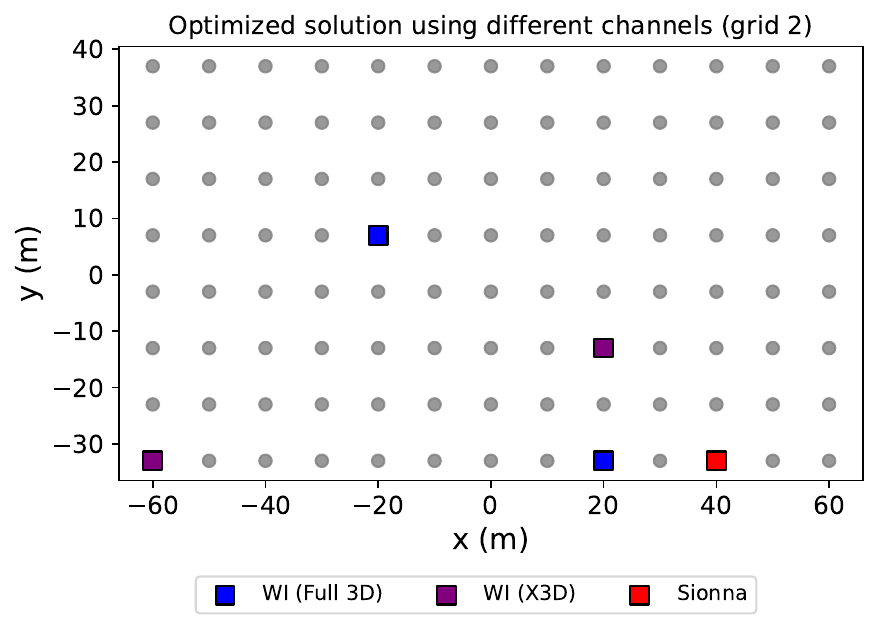}
    \label{fig:site_specific_results_grid2}
    }
    \caption{Optimized solution for \acp{GW} placement, in Etoile (using grid pattern 1) and Street Canyon (using grid pattern 2) scenarios, considering channels from Sionna \ac{RT} and \ac{WI} using X3D and Full 3D algorithms.}
    \label{fig:sionna_results}
\end{figure}

For the site-specific channels, we obtained the solutions depicted in Figs.~\ref{fig:site_specific_results_grid1} and \ref{fig:site_specific_results_grid2}, which show the position of each allocated \ac{GW} for the Etoile and Street Canyon scenarios, respectively, considering the channels created with Sionna \ac{RT} and \ac{WI}. The results obtained in Etoile using the Sionna \ac{RT} and \ac{WI} channels also yield an optimized solution that differs from that of the other site‑independent channels.

 Using the channels obtained with the \ac{WI} X3D \ac{RT} algorithm in the Etoile scenario, $3$ \acp{GW} were deployed, with one \ac{GW} placed at the same location as the \ac{GW} suggested when using the Sionna \ac{RT} channel. In contrast, when using the Full 3D algorithm, only $2$ \acp{GW} are suggested, and one of their positions coincides with those proposed by the optimization model for the \ac{WI} X3D channel. Finally, when using the Sionna channels, a larger number of \acp{GW} is suggested, equal to 9, highlighting a major difference between the algorithms employed in the two \ac{RT} simulators. For the Street Canyon, the variability is higher in comparison with the corresponding results from site-independent channels, with the simulations under \ac{WI} channels suggesting 2 \acp{GW} to meet the requirements.

\begin{figure}[!h]
    \centering
    \includegraphics[scale=0.54]{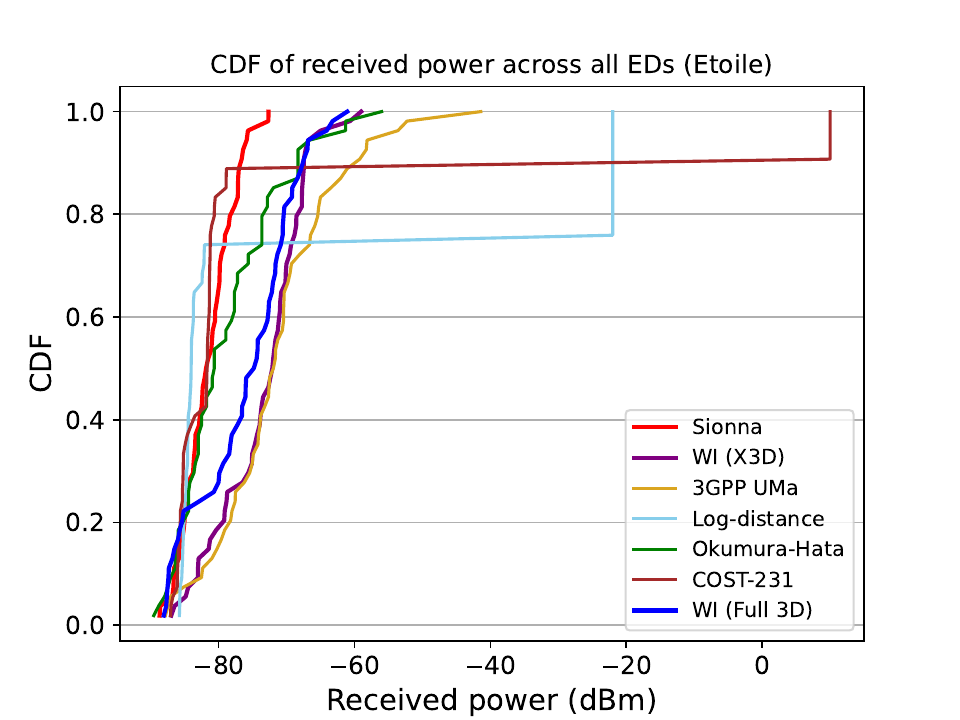}
    
    \caption{Received power \ac{CDF} under different channel models in grid pattern 1. It is possible to see the distance between the received power distribution that come from \ac{RT} channels in comparison to other site-independent models.}
    \label{fig:path_gain_cdf}
\end{figure}

In general terms, the difference between this number of \acp{GW} and the position coordinates, considering all channel models, is due to the distribution of the collected received power data. This difference can be seen in Fig.~\ref{fig:path_gain_cdf}, which shows the received power \ac{CDF} of each channel model, considering the solutions with a fixed threshold $\rho = -90$ \si{dBm}. From this figure, it can be observed that even with 9 \acp{GW} deployed under the Sionna \ac{RT} channel, the range of received power values for each \ac{ED} remains lower than the received power obtained when using a channel from \ac{WI}, with fewer \acp{GW} deployed in the Etoile scenario. Finally, the numerical results for both network grids, the number of suggested \acp{GW}, and the average and standard deviation \ac{ED} of received power for each channel model are described in Table~\ref{tab:power_threshold_scenarios}, also considering a fixed received power threshold of $\rho=-90$ \si{dBm}.

% \begin{table}[!h]
%     \centering
%     \caption{The corresponding number of suggested \acp{GW} and average \ac{ED} received power for each channel model}
%     \scalebox{1.1}{\begin{tabular}{lcc}
%     \toprule
%        Channel model  & \# Suggested \acp{GW} & Avg. \ac{ED} rec. power\\
%     \midrule
%          COST-231 & $11$ & $-83.11$ \\
%          Okumura-Hata & $1$ & $-78.65$\\
%          Log-distance  & $17$ & $-68.15$\\
%          3GPP-UMa & $1$ & $-71.31$\\
%          Sionna & 9 & $-81.47$\\
%          WI (X3D) & $3$ & $-73.39$\\
%          WI (Full 3D) & $2$ & $-75.99$\\
%     \bottomrule
%     \end{tabular}}
%     \label{tab: power_threshold}
% \end{table}

An additional analysis that can be made with the \acp{GW} positions determined after the optimization process, is to compare the difference in the received power for each \ac{ED} under different channel models. In particular, we take the received power under the \ac{WI} (X3D) model as the ground-truth and evaluate the received power differences for the same \ac{ED} when assigned to other channel models. Hence, Table~\ref{tab:relative_mse} describes the \ac{MSE} for each comparison. From these results, it is possible to observe that in a more crowded area (Etoile scenario with grid 1), the ED received power under 3GPP-UMa is the site-independent channel that most closely matches the received power obtained by the WI X3D model. In contrast, in the Street Canyon scenario with grid 2, the log-distance is the site-independent channel that more closely approximates the ED received power from \ac{WI} X3D. In this way, both behaviors are expected since, in low-density scenarios, received power is mainly driven by distance, which explains the good fit of the log-distance model. Conversely, in dense urban environments, where blockage, shadowing, and line-of-sight and non line-of-sight transitions dominate, the 3GPP-UMa model provides a better approximation due to its more detailed representation of these effects.

\begin{table}[!h]
    \centering
    \caption{Relative error in terms \ac{ED} received power considering the received power from \ac{WI} X3D model as ground-truth}
    \begin{tabular}{ccc}
    \toprule
       Channel model & MSE (Grid 1) & MSE (Grid 2) \\
       \midrule
        Okumura-Hata & $139.99$ & $477.84$\\ 
        COST-231 & $888.31$ & $142.41$ \\ 
        Log-distance & $825.51$ & $131.72$\\ 
        3GPP-UMa & $113.25$ & $244.30$\\ 
        Sionna & $118.12$ & $137.85$\\ 
        WI (Full 3D) & $\mathbf{52.99}$ & $\mathbf{54.93}$\\
        WI (X3D) & --- & ---\\ 
    \bottomrule
    \end{tabular}
    \label{tab:relative_mse}
\end{table}

\subsection{Parametric analysis with different thresholds}

\begin{figure}[!h]
    \centering
    \includegraphics[scale=0.52]{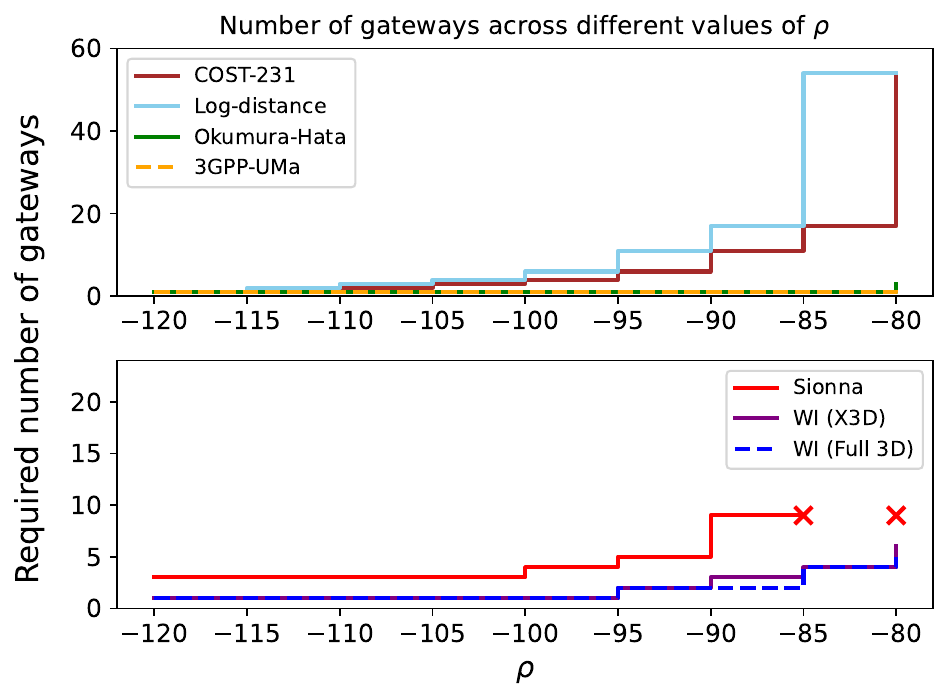}
    \caption{Number of \acp{GW} deployed considering different $\rho$ values in grid pattern 1. In this plot, we compare the number of suggested \acp{GW} considering site-independent channels (upper) and the site-specific channels (bottom). The ``$\times$'' mark represents an infeasible solution for $\rho=$-85 and $\rho=$-80 dBm.}
    \label{fig:across_rho}
\end{figure}

In the second set of evaluations, we vary the value of $\rho$, and it is possible to verify the optimization model from a general perspective in terms of the number of suggested \acp{GW} in all grid organizations. In the grid pattern 2, we found fewer variability in terms of the number of \ac{GW} required to meet the constraints across different values of $\rho$. In this case, 1 \ac{GW} is enough when using all options of site-independent channels, and 1 or 2 is enough when a site-specific channel is used. However, when in consider a more complex scenario provided by grid pattern 1, we have some important insights in the context of network planning. In this sense, Fig.~\ref{fig:across_rho} shows the number of \acp{GW} suggested for each model across different values of $\rho$, considering an interval that ranges from $-120$ to $-80$ \si{dBm}. In this figure, we can also observe the odd behavior of the optimization model under the COST-231 and log-distance channel models. This behavior starts at $\rho = -85$ for log-distance and $\rho = -80$ for COST-231, with the optimization model proposing at least $17$ times more \acp{GW} than models such as Okumura–Hata, resulting in a total of 54 \acp{GW}. This indicates that, in the context of network planning, some of the solutions suggested under certain channel models can be highly inaccurate, even with a certain similarity in terms of received power \ac{CDF}, as seen in Fig.~\ref{fig:path_gain_cdf}. In the case of the COST-231 model, this results in an unrealistic solution, which is explained by the fact that the COST-231 model was originally designed for higher-frequency urban macrocell scenarios. Hence, the predicted path loss becomes overly pessimistic in LoRaWAN bands and environments. Combined with strict receiver sensitivity thresholds, this leads to artificial coverage fragmentation, where the optimization degenerates to placing one \ac{GW} per \ac{ED}.

\begin{figure}[!h]
    \centering
    \includegraphics[scale=0.33]{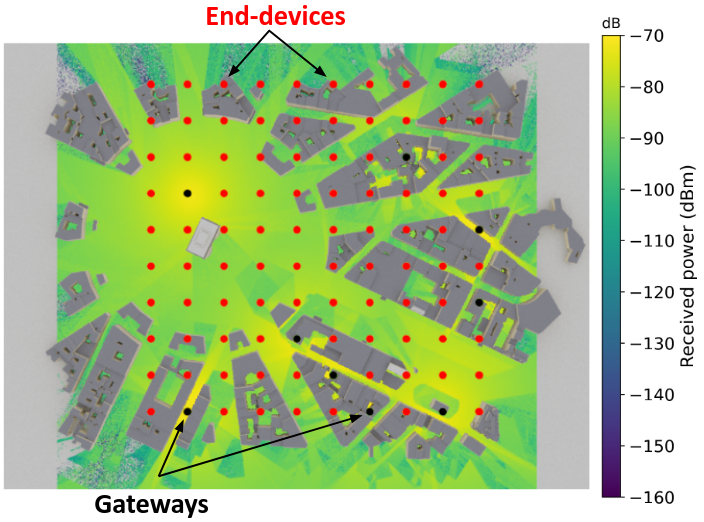}
    \caption{Coverage map from the optimized configuration considering \ac{RT} channels from Sionna \ac{RT} in Etoile scenario. From all possible positions represented by a red dot, the black dots represent the chosen \ac{GW} positions.}
    \label{fig:optimized_solution_sionna}
\end{figure}

Furthermore, for the Sionna \ac{RT} channels, the infeasibility of finding an optimized solution for certain values of $\rho$ can also be explained by the geometry of the scenario in which the received power is collected and the \ac{RT} algorithm used by the simulator~\cite{Zhu2024}. We can observe that even close to the \ac{GW}, the \ac{ED} coverage may be attenuated by scattering. This is more evident when we visualize a 3D representation of the scenario, considering the 9 chosen \ac{GW} positions. For the obtained channels with Sionna \ac{RT}, we can observe that even in close proximity to the \ac{GW}, \ac{ED} coverage can be significantly attenuated due to scattering, as we can visualize in the 3D representation of the scenario with the coverage map, as depicted in Fig.~\ref{fig:optimized_solution_sionna}.

\subsection{High-level metrics performance analysis}
\begin{figure}[!h]
    \centering
    \includegraphics[scale=0.52]{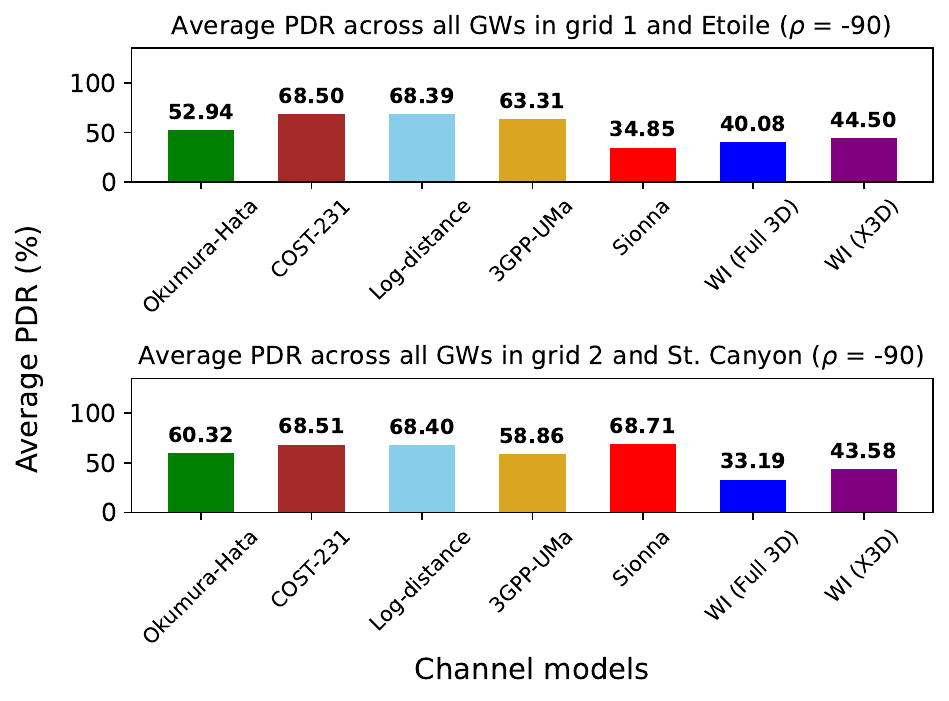}
    \caption{Evaluated average \ac{PDR} considering site-independent and site-specific channels across different network scenarios and grid patterns.}
    \label{fig:avgpdr_per_channel}
\end{figure}

In the third set of evaluations, to evaluate the impact of \ac{GW} placement from a high-level performance perspective, we reuse the ns-3 LoRaWAN environment introduced in step (3) of Fig.~\ref{fig:rt_ns3_lorawan} and assess the \ac{PDR} with the optimized \ac{GW} positions for a sensitivity threshold of $\rho = -90$\,\si{dBm} across 10 realizations. For environments with grids 1 and 2, we assume that each \ac{ED} transmits 1000 and 1040 packets, respectively, to the \acp{GW}, using \ac{SF} 7 and a simulation time of 600 seconds per channel. This configuration results in an airtime of $T_{\text{airtime}} = 0.1$\,\si{s} and a packet rate of $\lambda = 1.67$ \si{packets/s}, resulting in an offered traffic of $L=0.167$. Therefore, for each channel model (in both grid patterns), we have the average \ac{PDR} across all \acp{GW}, depicted in Fig.~\ref{fig:avgpdr_per_channel}, and numerically described in Table~\ref{tab:pdr_time_results} with a confidence interval of 95\%.

\begin{table*}[!h]
    \centering
    \caption{Average and standard deviation values from simulation time (minutes) and \ac{PDR} (\%) across all chosen \acp{GW} in different network environments}
    \scalebox{0.88}{
    \begin{tabular}{lcccccccccc}
    \toprule
    & \multicolumn{5}{c}{Grid 1} & \multicolumn{5}{c}{Grid 2} \\
    \cmidrule(lr){2-6} \cmidrule(lr){7-11}
    Channel model 
    & Avg. PDR & Std. PDR & PDR 95\% CI & \makecell{Avg.\\ simul. time} & \makecell{Std.\\simul. time}  
    & Avg. PDR & Std. PDR & PDR 95\% CI & \makecell{Avg.\\ simul. time} & \makecell{Std.\\simul. time} \\
    \midrule
    Okumura-Hata  & $52.94$  & $0.29$ & $52.94 \pm 0.21$ & $0.01$ & $3\times10^{-5}$ & $60.32$& $3.29$ & $60.32 \pm 2.35$ & $0.01$ & $4\times10^{-5}$ \\
    COST-231 & $68.50$ & $0.08$ & $68.50 \pm 0.06$ & $0.01$ & $2 \times 10^{-5}$& $68.51$ & $0.44$ & $68.51 \pm 0.31$ & $0.01$ & $5\times10^{-5}$ \\
    Log-distance & $68.39$ & $0.02$ & $68.39 \pm 0.01$ & $0.01$ & $3\times 10^{-5}$ & $68.40$ & $0.02$ & $68.40 \pm 0.02$ & $0.01$ & $3\times10^{-5}$ \\
    3GPP-UMa & $63.31$  & $1.26$ & $63.31 \pm 0.9$ & $0.01$ & $4\times10^{-5}$ & $58.86$ & $3.69$ & $58.86 \pm 2.64$ & $0.01$ & $5\times10^{-5}$\\
    Sionna        & $34.85$  & $0$ & $34.85$ & $0.27$ & $0.02$ & $68.71$ & $0$ & $68.71$ & $0.17$ & $0.01$ \\
    WI (Full 3D)  & $40.08$  & $0$ & $40.08$ & 61.65 & $0.06$ & $33.19$ & $0$ & $33.19$ & $2.69$ & $0.04$ \\
    WI (X3D) & $44.50$  & $0$ & $44.50$ & $24.95$ & $0.08$ & $43.58$ & $0$ & $43.58$ & $4.37$ & $0.02$ \\
    \bottomrule
    \end{tabular}}
    \label{tab:pdr_time_results}
\end{table*}

The average \ac{PDR} shown in Fig.~\ref{fig:avgpdr_per_channel} highlights the impact of channel modeling on high-level metrics for network planning. For site-specific channel models in the Etoile scenario, the results are pessimistic, as these models explicitly account for scattering effects in the simulations. This leads to increased attenuation of the transmitted signal from the \ac{ED} to the \ac{GW}, resulting in an average \ac{PDR} below $45\%$. In contrast, site-independent channel models yield more optimistic results for network planning, with each model achieving an average \ac{PDR} above $52\%$. For the experiments using grid pattern 2, the optimistic results also remain when we use the site-independent channels. However, the first difference is related to the use of site-specific channels from Sionna, which, in this case, presented an average \ac{PDR} similar to the site-independent channels. This similarity can be attributed to the level of detail implemented in the ray tracer. As a consequence, in relatively simple scenarios, with fewer buildings and scattering events, such as the Street Canyon case, the path gain results from the Sionna \ac{RT} model closely resemble those obtained from site-independent channel models. For the \ac{PDR} using an environment with \ac{WI} channels, the performance remains similar. Overall, these results immediately suggest potential hidden problems in network planning, since the choice to use site-independent channel models may conceal serious issues in network environments close to the real-world. Thereby, the use of more realistic \ac{RT}-based channels can show that \ac{LoRaWAN} networks can lose more than half of their packets during communication. This observation reinforces the need to perform network simulations that account for the specific environment in order to improve the fidelity of performance results in the early phases of network deployment, intended to adopt complementary multi-objective optimization considering other parameters of the network, such as different \acp{SF}, energy metrics, and transmission power.

\subsection{Computational cost analysis}
\begin{figure}[!h]
    \centering
    \includegraphics[scale=0.5]{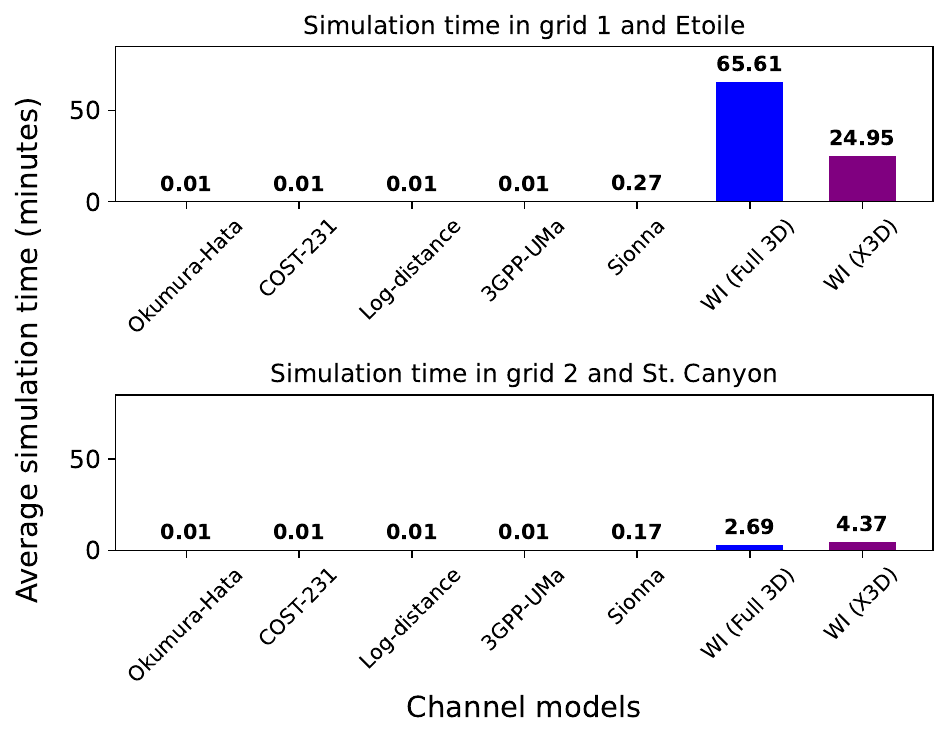}
    \caption{Simulation time, in minutes, to obtain physical layer metrics across different channel models and scenarios.}
    \label{fig:simulation_time}
\end{figure}

\begin{figure}[!h]
    \centering
    \includegraphics[scale=0.62]{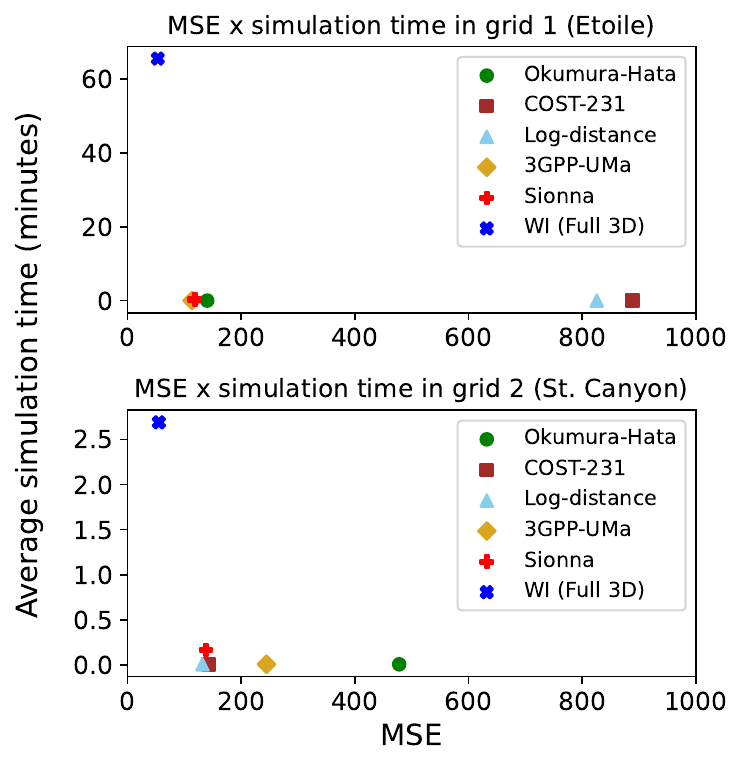}
    \caption{Tradeoff between average simulation time and \ac{MSE} considering the \ac{ED} received power from \ac{WI} X3D as the ground truth.}
    \label{fig:tradeoff_error_time}
\end{figure}

Finally, among the aforementioned solutions, it is possible to infer a clear trade-off between the fidelity of the results and the computational cost required to obtain the metrics used in the optimization model. Accordingly, Fig.~\ref{fig:simulation_time} and Table~\ref{tab:pdr_time_results} report the average simulation time across 10 realizations, in minutes, needed to compute the physical-level metrics used in the optimization model, using both grid patterns and 3D scenarios when a site-specific channel is employed. This figure reveals two main discrepancies. First, there is a much larger difference in simulation time between the site-independent and site-specific models, with the former requiring less than one second to obtain the physical level metrics. The second discrepancy is the striking difference between the two \ac{RT} simulators: the simulation time using the Etoile scenario in the \ac{WI} is at least 92 times higher than that of Sionna \ac{RT}. This gap can become even larger if the Full 3D algorithm is used in the \ac{WI}. In this sense, all these site-specific \ac{RT} simulations took place in 3D scenarios with fewer than $20\,000$ faces. Even with this number of polygons representing the objects in the scenario, high‑detail ray tracers such as \ac{WI} still incur a high computational cost to produce accurate metrics, such as path gain. This trade-off becomes clearer when examining a scatter plot, with the simulation time on the $y$-axis and the \ac{MSE} values from Table~\ref{tab:relative_mse} on the $x$-axis, as depicted in Fig.~\ref{fig:tradeoff_error_time}. 

From these results, we can observe that achieving the lowest \ac{MSE} with \ac{WI} channels evaluated using the Full 3D algorithm requires a large simulation time. Moreover, for some site-specific and site-independent channels, the trade-off between simulation time and \ac{MSE} is comparable. For example, in environments with grid~1 and Etoile, the trade-off for Okumura-Hata, 3GPP-UMa, and Sionna is similar. Likewise, in environments with grid~2 and Street Canyon, this similarity is observed for Log-distance, COST-231, and Sionna. Therefore, this final result underscores the sensitivity of network planning in this context. In the first evaluation, we observed that even with similar \ac{ED} received power \ac{CDF}, it stemmed from different spatial organizations of \acp{GW}. Now, we see that different \ac{PDR} outcomes analyzed in the previous subsection come from channel models with similar tradeoffs between simulation time and \ac{MSE}.

\section{CONCLUSIONS AND FUTURE WORKS}
\label{sec:conclusion} 
In this work, we proposed a framework that integrates \ac{RT} tools, such as Sionna and \ac{WI}, with the discrete-event network simulator ns-3 to support network planning for the \ac{GW} placement problem in different LoRaWAN scenarios. Using this integrated framework, we assess how different channel models influence the outcomes of the optimization model. In this sense, we show that the quality of the network planning solution is strongly dependent on the channel type when a site-independent approach is used, and it can also vary with the choice of \ac{RT} simulator when a site-specific approach is adopted. This choice influences the number of \acp{GW} that the solution suggests deploying and the position of each one. Numerically, for site-independent models such as log-distance, this discrepancy can be highly inaccurate, reaching values up to $17$ times greater than those suggested by models such as Okumura–Hata and 3GPP-UMa. For the site-specific channels, certain coverage-related received-power constraints admit no feasible solution (e.g., when using Sionna \ac{RT}). However, for the same parameter values, feasible solutions do exist when using \ac{WI} channels, also varying the \ac{RT} algorithms. This behavior highlights the importance and the difference between two \ac{RT} simulators and their impact on network planning. 

Furthermore, we show the impact of the \ac{GW} positioning, considering high-level metrics such as \ac{PDR} in different channel models. In this sense, the results lead in the same direction when we only consider the impact on the received power: a decisive impact when adopting a site-specific channel model instead of a site-independent one, with the former models yielding a more pessimistic \ac{PDR} compared to the latter. Finally, we illustrate that all these results involve a trade-off between the computational cost and the solution accuracy required to obtain them.

For future work, we aim to enhance our proposed environment by incorporating additional metrics that capture the relationship between the \ac{GW} and the \ac{ED}. In the short term, we will focus on improving \ac{LoRaWAN} network simulations to consider capacity and energy constraints. This improved environment will be used to testing different transmission power levels and \ac{SF} configurations, essential for mitigating collision problems in transmission. In the long term, we aim to integrate our proposed ns-3 \ac{LoRaWAN} framework into an \ac{NDT} platform that supports comprehensive \textit{what-if} analysis. In this case, since an \ac{NDT} inherently requires a physical counterpart, this future task will help bridge the comparison gap not addressed in this work between our proposed site-specific optimization approach and real-world data obtained from measurement campaigns. Finally, with this \ac{NDT} deployed, it will enable closed-loop applications such as optimizing \ac{GW} placement prior to real-world deployment, in a manner similar to existing work on \ac{5G} transport networks~\cite{abenathar2025}.

\bibliographystyle{IEEEtran}
\bibliography{references}
\begin{IEEEbiography}
[{\includegraphics[width=1in,height=1.15in,clip]{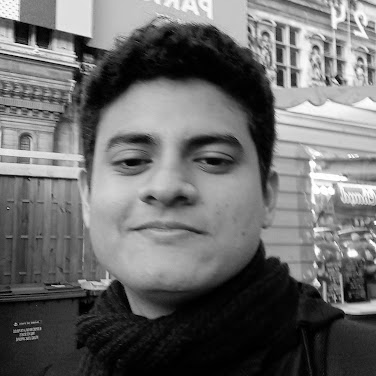}}]{Cláudio Modesto }  received his B.Sc. degree in computer engineering in 2024 and his M.Sc. degree in electrical engineering with emphasis on telecommunications from the Electrical Engineering Graduate Program, UFPA, in 2025, where he is currently pursuing the
Ph.D. degree. Since 2023, he has been a member of the Research and Development Center for Telecommunications, Automation, and Electronics (LASSE). His research interests include signal processing for wireless communications, as well as machine learning and deep learning applied to graph-structured data.
\end{IEEEbiography}
\begin{IEEEbiography}[{\includegraphics[width=1in,height=1.1in]{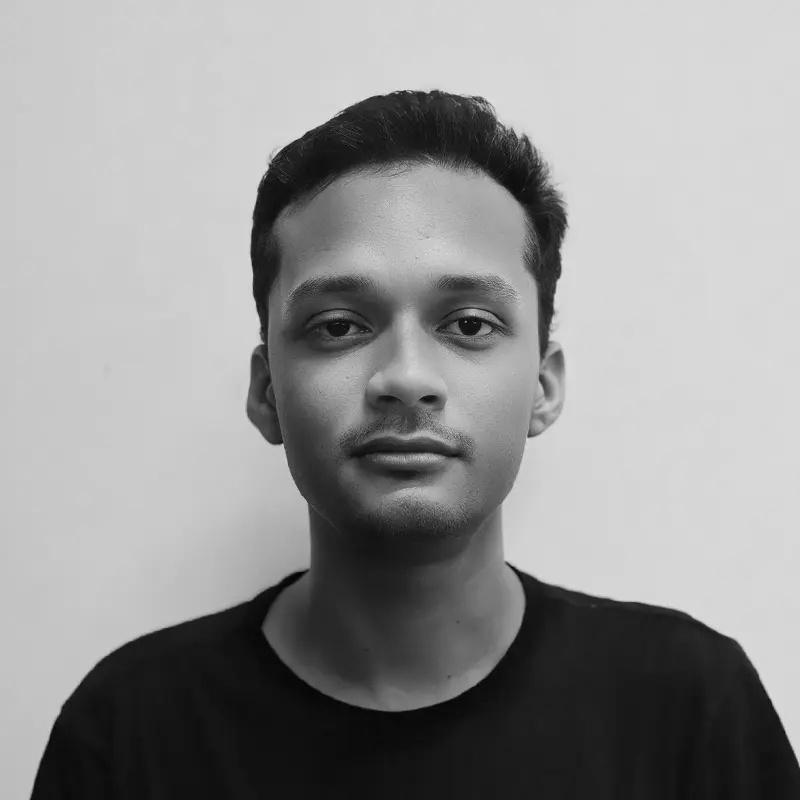}}]{Lucas Mozart } received his B.Sc. degree in computer engineering from the Federal University of Pará (UFPA), Belém, Pará, Brazil, in 2025. He is currently pursuing an M.Sc. degree in Electrical Engineering at UFPA. He has been part of the Research and Development Center for Telecommunications, Automation, and Electronics (LASSE) since 2023. His current research interests include signal processing and machine learning applied to wireless communications.
\end{IEEEbiography}
\begin{IEEEbiography}[{\includegraphics[width=1in,height=1.25in,clip,keepaspectratio]{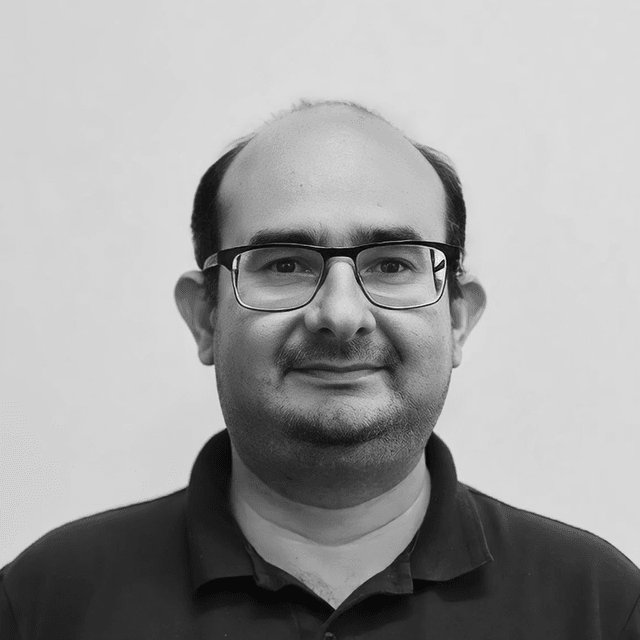}}]{Glauco Gonçalves } has been an associate professor at the Federal University of Pará since 2021. He was formerly a professor at the
Federal Rural University of Pernambuco (2013-2021) where he worked developing research and teaching. He has a Ph.D (2012) and a master’s degree (2007) in computer science from the Informatics Center of the Federal University of Pernambuco. Glauco received his bachelor’s degree in computer science from the Federal University of Pará in 2005. He works in the area of computer science applying computational modeling methods in different areas such as 6G networks, cloud computing, and Internet of Things.
\end{IEEEbiography}
\begin{IEEEbiography}[{\includegraphics[width=1in,height=1.25in,clip,keepaspectratio]{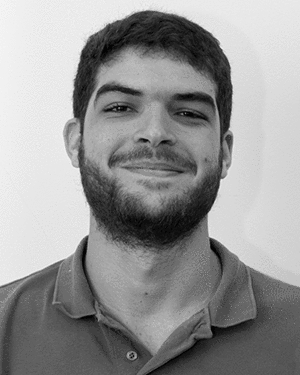}}]{Cleverson Nahum } received the B.Sc. degree in computer engineering, and the M.Sc. and Ph.D degrees in electrical engineering (telecommunications) from the Federal University of Pará, Belém,
Brazil, in 2019, 2021, and 2024, respectively. He is
currently a postdoctoral researcher in the SAMURAI FAPESP/SP Project, working with the Research and Development Center for Telecommunications, Automation, and Electronics (LASSE). His research interests include machine learning for wireless communications, OpenRAN, and intent-based networks.
\end{IEEEbiography}
\begin{IEEEbiography}[{\includegraphics[width=1in,height=1.25in,clip,keepaspectratio]{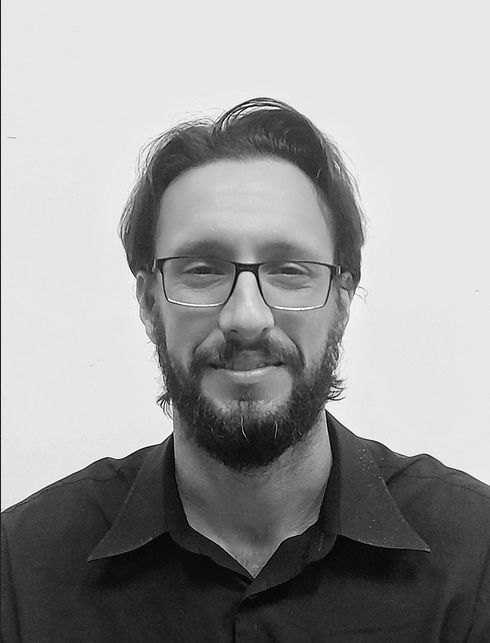}}]{Bruno Castro } Bachelor's degree in Telecommunications Engineering from the Institute of Higher Studies of the Amazon (2006), a Master's in the Graduate Program in Electrical Engineering from the Federal University of Pará (2010), and a Ph.D. in the Graduate Program in Electrical Engineering from the Federal University of Pará (2014). Currently, an Adjunct Professor at the Federal University of Pará. Has experience in the field of Electrical Engineering, with an emphasis on mobile communications, working primarily on the following topics: signal propagation, digital transmission, wireless networks, quality of experience (QoE), cross-layer modeling, 5G Networks, internet of things (IoT), and data science.
\end{IEEEbiography}
\begin{IEEEbiography}[{\includegraphics[width=1in,height=1.25in,clip,keepaspectratio]{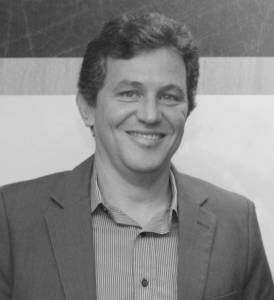}}]{Aldebaro Klautau } (Senior Member, IEEE)
received a bachelor's degree (Federal University of Pará, UFPA, 1990), M.Sc. (Federal Univ. of Santa Catarina, UFSC, 1993) and Ph.D. degrees (University of California at San Diego, UCSD, 2003) in electrical engineering. He is a full professor at UFPA, where he is the ITU Focal Point and coordinates the LASSE Research Group. He is a researcher at CNPq, Brazil, and is a senior member of the IEEE and the Brazilian Telecommunications Society (SBrT). His work focuses on machine learning and signal processing for communication.
\end{IEEEbiography}

\end{document}